\documentstyle[aps,epsf]{revtex}
\def\abstract#1{\vskip 7mm 
        \begin{center}{\large Abstract}\par \smallskip
                \begin{minipage}[c]{12cm}
                        \small #1
                \end{minipage}
        \end{center}}
\def\title#1{\begin{center}{\Large\bf #1}\end{center}}
\def\author#1{\vskip 5mm \begin{center}{#1}\end{center}}
\def\address#1{\begin{center}{\it #1}\end{center}}
\makeatletter
\@ifundefined{lesssim}{}{}
\@ifundefined{gtrsim}{}{}
\def\vereq#1#2{\lower3pt\vbox{\baselineskip1.5pt \lineskip1.5pt
\ialign{$\m@th#1\hfill##\hfil$\crcr#2\crcr\sim\crcr}}}
\makeatother
\begin{document}
\def\d{d}
\def\p{\partial}
\def\w{\wedge}
\def\o{\otimes}
\def\half{{\scriptstyle \frac{1}{2}}}

\title{%
  Black holes in three-dimensional
  Einstein-Born-Infeld-dilaton theory
  \smallskip \\
}
\author{%
  Ryo Yamazaki \footnote{E-mail:yamazaki@tap.scphys.kyoto-u.ac.jp}
  and Daisuke Ida\footnote{E-mail:ida@tap.scphys.kyoto-u.ac.jp}
}
\address{%
  Department of Physics, Kyoto University, Kyoto 606-8502, Japan
}
\abstract{
The three-dimensional static and circularly symmetric solution of the 
Einstein-Born-Infeld-dilaton system is derived.
The solutions corresponding to low energy string theory
are investigated in detail, which include black hole solutions
if the cosmological constant is negative and the mass parameter
exceeds a certain critical value.
Some differences between the Born-Infeld nonlinear electrodynamics and
the Maxwell electrodynamics are revealed.
}

\section{Introduction}
The black hole is one of the fundamental objects in gravity theories.
What we shall consider here is a black hole in the 
Einstein-Born-Infeld-dilaton theory motivated by string theory.
We have a clear picture of an astrophysical black hole as a final
product of the gravitational collapse owing to the uniqueness theorem 
of the stationary asymptotically flat black hole solution.
On the other hand, there are various kinds of black objects the string
theory since there emerge many matter fields as in the present case.
It seems hard to reveal the general properties of such stringy black holes.
We want to approach this problem.

The Born-Infeld nonlinear electrodynamics has attracted much
attention in the context of string theory.
It was originally introduced for
the purpose of solving various problems of divergence
appearing in the Maxwell theory \cite{BI}.
It has later been shown that the Born-Infeld theory naturally arises
in the low energy limit of the open string theory \cite{FT,ACNY}.

In the open and the closed bosonic string theory,
there are four massless states of strings;
the dilaton field $\phi$, the $U(1)$ gauge field $A_i$,
the gravitational field $g_{ij}$, 
and the Kolb-Ramond antisymmetric tensor field $B_{ij}$.
When one considers the Born-Infeld electrodynamics motivated by
the string theory, he might include other massless fields together
with the electromagnetic field.
In this paper, 
we however neglect $B_{ij}$ for simplicity.
Then, the Einstein-Born-Infeld-dilaton system describes interactions
among massless fields $\phi$, $A_i$, and $g_{ij}$.

Recently, numerical studies of the Einstein-Born-Infeld-dilaton system
have been done in four-dimensional static and spherically symmetric
space-time \cite{TT,TT2,ClGa}.
Here we turn to the three-dimensional case, and show that
the analytic solution describing the black hole can be obtained.
The three-dimensional gravity is a simple version of general relativity
and gives one useful approach to the black hole 
physics \cite{Carlip,BTZ,BHTZ}.
Three-dimensional black holes also arise in some higher dimensional
theories \cite{Hyun}.
We may anyway get an insight into the property of the Born-Infeld
theory by investigating the three-dimensional black hole as the
simplest example.

This paper is organized as follows.
In Sec.~\ref{BasicEq}, we derive basic equations of the
Einstein-Born-Infeld-dilaton system for general dilaton couplings,
and give an analytic solution under appropriate assumptions.
In Sec.~\ref{StringCase}, we analyze in detail 
the solution to the low energy string thory.
We especially focus on the effects of the nonlinearity of the
Born-Infeld field, where some differences between the Maxwell 
and the Born-Infeld fields are revealed.
In Sec.~\ref{Conclusion}, we summarize the results.

\section{Basic Equations and Solutions}
\label{BasicEq}
We first consider general form of Lagrangian which describes
nonlinear electrodynamics.
We adopt the action which is written in string frame as
\begin{equation}
S=\int\d^3 x\sqrt{-g}e^{-2\phi}
\left( R-2\Lambda+4(\partial\phi)^2+{\cal L}[F^2]\right),
\label{SframeAction}
\end{equation}
where $F^2=F_{ij}F^{ij}$ and ${\cal L}$ is its functional.
The action in Einstein frame is given via the conformal transformation
$g_{ij}\mapsto e^{4\phi}g_{ij}$ by
\begin{equation}
S=\int\d^3 x\sqrt{-g}\left( R-2e^{4\phi}\Lambda
-4(\partial\phi)^2+e^{4\phi}{\cal L}[e^{-8\phi}F^2]\right).
\end{equation}
We consider an action with general dilaton coupling constants $\alpha$,
$\beta$ and $\gamma>0$;
\begin{equation}
S=\int\d^3 x\sqrt{-g}\left( R-2e^{\alpha\phi} \Lambda
-\gamma (\partial\phi)^2+e^{\beta\phi}{\cal L}[e^{-2\beta\phi}F^2]\right).
\label{action}
\end{equation}
The equations of motion of the dilaton field $\phi$ and the electromagnetic
field $F$ are
\begin{equation}
2\gamma\phi_{,i}{}^{;i}-2\alpha\Lambda e^{\alpha\phi}+\beta e^{\beta\phi}
{\cal L}-2\beta e^{-\beta\phi}F^2 \dot{{\cal L}}=0,\\
\label{dilaton}
\end{equation}
\begin{equation}
\left(e^{-\beta\phi}\dot{{\cal L}}F^{ij}\right)_{;j}=0,
\label{elemag}
\end{equation}
respectively, where $\dot{{\cal L}}[x]=\delta {\cal L}/\delta x$.
The Einstein equation is
\begin{eqnarray}
R_{ij}=4\phi_{,i}\phi_{,j}+\left(2e^{\alpha\phi}\Lambda
-e^{\beta\phi}{\cal L}\right)g_{ij}
{}-2e^{-\beta\phi}\dot{{\cal L}}(F_{ik}F_j{}^k-F^2g_{ij}).
\label{ricci}
\end{eqnarray}

We consider the Born-Infeld nonlinear electrodynamics, which corresponds to
\begin{equation}
{\cal L}[x]=4b^2\left[1-\left(1+\frac{x}{2b^2}\right)^{\half}\right],
\label{BI_Lag}
\end{equation}
where $b>0$ is the Born-Infeld parameter.
In the limit of large $b$, this gives
Maxwell Lagrangian;
${\cal L}\rightarrow -x$.

The three-dimensional static and circularly symmetric space-time can be
written in the form
\begin{equation}
g=-f(r)\d t^2+\frac{e^{2\delta(r)}}{f(r)}\d r^2+r^2\d\varphi^2.
\label{metric}
\end{equation}
The electromagnetic field is assumed to have the following pure
electrostatic form
\begin{equation}
F=e^{\beta\phi(r)+\delta(r)}E(r)\d t\wedge\d r.
\label{strength}
\end{equation}
Then, Eq.~(\ref{elemag}) becomes
\begin{equation}
\left(\frac{rE}{\sqrt{1-E^2/b^2}}\right)'=0,
\label{bimaxwell}
\end{equation}
where prime denotes derivative with respect to $r$.
This can be integrated to give
\begin{equation}
E=h^{-\half}\frac{Q}{r},
\label{elec}
\end{equation}
where
\begin{equation}
h=1+\frac{Q^2}{b^2r^2},
\end{equation}
and $Q$ is an integration constant.
The constant $Q$ is the electric charge, namely
\begin{equation}
Q=-{1\over 4\pi}\oint_{\Gamma}e^{-\beta\phi}\dot{{\cal L}}*F,
\end{equation}
for any smooth closed spacelike curve $\Gamma$ enclosing $r=0$.
The equation of motion of the dilaton field
(\ref{dilaton}) becomes
\begin{equation}
\frac{\gamma e^{-\delta}}{r}(rfe^{-\delta}\phi')'-\alpha\Lambda
e^{\alpha\phi}+2\beta b^2e^{\beta\phi}\left(1-h^{\half}\right)=0.
\label{bidilaton}
\end{equation}
The Einstein equations are
\begin{eqnarray}
\left(f''+\frac{f'}{r}-f'\delta'\right)e^{-2\delta}
=-4\Lambda e^{\alpha\phi}+8b^2e^{\beta\phi}
\left(1-h^{\half}+\frac{Q^2}{2b^2r^2}h^{-\half}\right),
\label{ricci00}
\end{eqnarray}
\begin{equation}
\delta'=\gamma r \left(\phi'\right)^2,
\label{ricci00-11}
\end{equation}
\begin{equation}
\frac{f'-f\delta'}{r}e^{-2\delta}
=-2\Lambda e^{\alpha\phi}+4b^2e^{\beta\phi}(1-h^{\half}).
\label{ricci22}
\end{equation}
Here, we assume the following form of the metric function $\delta$;
\begin{equation}
\delta=n \ln\left( \frac{r}{r_0}\right),~~~{\rm for}~~r>0,
\label{delta}
\end{equation}
where $n$ and $r_0>0$ are constants. From Eq.~(\ref{ricci00-11}),
it turns out that $n$ must be positive, and then the dilaton field
$\phi$ becomes
\begin{equation}
\phi=\pm\left(\frac{n}{\gamma}\right)^{\half}
\ln\left( \frac{r}{r_1}\right),
\label{phi}
\end{equation}
with a positive constant $r_1$.
Then, Eqs.~(\ref{bidilaton}) and (\ref{ricci22}) become
\begin{eqnarray}
\left(\frac{f}{r^n}\right)'=\pm\frac{r^{n+1}}
{(\gamma n)^{\half}r_0{}^{2n}}
\left[\alpha\Lambda\left(\frac{r}{r_1}\right)^{\pm\alpha\sqrt{n/\gamma}}
-2\beta b^2\left(\frac{r}{r_1}\right)^{\pm\beta\sqrt{n/\gamma}}
(1-h^{\half})\right],
\label{f1}
\end{eqnarray}
and
\begin{eqnarray}
\left(\frac{f}{r^n}\right)'=\frac{r^{n+1}}{r_0{}^{2n}}
\left[-2\Lambda\left(\frac{r}{r_1}\right)^{\pm\alpha\sqrt{n/\gamma}}
+4 b^2\left(\frac{r}{r_1}\right)^{\pm\beta\sqrt{n/\gamma}}(1-h^{\half})
\right],
\label{f2}
\end{eqnarray}
respectively.
These two equations (\ref{f1}) and (\ref{f2}) are consistent if and only if
\begin{equation}
\alpha=\beta=\mp \sqrt{4n\gamma}.
\label{condition}
\end{equation}
Equation (\ref{condition}) includes the string case
($n=1$, $\alpha=\beta=\gamma=4$).
When the condition (\ref{condition}) is satisfied, Eqs.~(\ref{f1}) and
(\ref{f2}) become
\begin{equation}
\left(\frac{f}{r^n}\right)'=\left(\frac{r_1}{r_0}\right)^{2n}
\left[(4b^2-2\Lambda)-4b^2h^{\half}\right]r^{1-n}.
\label{f3}
\end{equation}
The solutions of Eq.~(\ref{f3}) are
\begin{equation}
f=\left(\frac{r_1}{r_0}\right)^{2n}\left[
\frac{4b^2-2\Lambda}{2-n}r^2-4b^2r^n\int^rr^{1-n}\left(1+\frac{Q^2}{b^2r^2}
\right)^{\half}
\d r\right],
\label{solution}
\end{equation}
for $n\ne 2$, and
\begin{equation}
f=\left(\frac{r_1}{r_0}\right)^4\left[
(4b^2-2\Lambda)r^2\ln \left(\frac{r}{r_2}\right)-4b^2r^2
\int^rr^{-1}\left(1+\frac{Q^2}{b^2r^2}\right)^{\half}\d r\right],
\label{f_n=2}
\end{equation}
for $n=2$.

The above solutions include known solutions as special cases.
When $n=0$, the dilaton field vanishes, and then 
Eq.~(\ref{solution}) becomes
\begin{equation}
f=-M-(\Lambda-2b^2)r^2-2b^2r\sqrt{r^2+Q^2/b^2}
-2Q^2\ln\left(r+\sqrt{r^2+Q^2/b^2}\right),
\end{equation}
which has been given by Cataldo and Garc\'{\i}a \cite{CG}.
In the limit of $b\rightarrow\infty$, the Born-Infeld field reduces
just to the Maxwell field.
Then Eq.~(\ref{solution}) gives
\begin{equation}
f=\left(\frac{r_1}{r_0}\right)^{2n}\left[
-Ar^n-\frac{2\Lambda}{2-n}r^2+\frac{2Q^2}{n}\right],
\label{CMa}
\end{equation}
for $n\ne 0,2$,
\begin{equation}
f=\left(\frac{r_1}{r_0}\right)^4\left[
-Ar^2-2\Lambda r^2\ln \left(\frac{r}{r_2}\right) +Q^2 \right],
\label{CMb}
\end{equation}
for $n=2$, and
\begin{equation}
f=-M-\Lambda r^2-2Q^2\ln\left(\frac{r}{r_3}\right),
\label{BTZQ}
\end{equation}
for $n=0$.
The equation (\ref{CMa}) coincides with the solution obtined 
by Chan and Mann \cite{CM},
and Eq.~(\ref{BTZQ}) corresponds to the charged BTZ solution \cite{BTZ}.
Other solutions can be obtained by taking special values of parameters
in Eqs.~(\ref{solution}) or (\ref{f_n=2}).
The Table \ref{class} shows the classification of solutions
found previously \cite{BTZ,BHTZ,CG,CM,MNY,MSW,SKL,BBL,DM,GSA}.
\begin{table}[htbp]
\begin{center}
\begin{tabular}{c c c c c c c c}
& $b$ & $\Lambda$ & $\alpha$, $\beta$, $\gamma$ & $n$
& $M$ & $Q$ & System \\ \hline\hline
Eqs.~(\ref{solution}), (\ref{f_n=2}) &&&
\multicolumn{2}{c}{$\alpha =\beta =\mp 2\sqrt{n\gamma}$}
&&& EBID$\Lambda$ \\ \hline
Eq.~(\ref{f_n=1}) &&& $\alpha =\beta =\gamma =4$ & $1$
&&& EBID$\Lambda$ (String) \\ \hline
Cataldo \& Garc\'{\i}a (1999) &&& $\alpha =\beta =0$ & $0$ 
&&& EBI$\Lambda$ \\ \hline
Chan \& Mann (1994) & $+\infty$ && 
\multicolumn{2}{c}{$\alpha =\beta =\mp 2\sqrt{n\gamma}\ \ , \ \ n\ne 2$}
&&& EMD$\Lambda$ \\ \hline
McGuigan et al. (1992) & $+\infty$ && $\alpha =\beta =\gamma =4$ & $1$ 
&&& EMD$\Lambda$ (String) \\ \hline
Mandal et al. (1991) & $+\infty$ && $\alpha =\beta =\gamma =4$ & $1$ 
&& $0$ & ED$\Lambda$ (String) \\ \hline
Sa et al. (1996) & $+\infty$ & $<0$ &
\multicolumn{2}{c}{$\alpha =\beta =n\gamma =4 \ \ , \ \ n\ne 1,2$}
&& $0$ & ED$\Lambda$ \\ \hline
Sa et al. (1996) & $+\infty$ & $<0$ & $\alpha =\beta =4,\gamma =2$ & $2$
& $0$ & $0$ & ED$\Lambda$ \\ \hline
Barrow et al. (1986) & $+\infty$ & $<0$ & $\alpha =-\sqrt{2} \ \ , \ \ 
\gamma =1$ & $1/2$ & $0$ & $0$ & ED$\Lambda$ \\ \hline
Ba\~{n}ados et al. (1992) & $+\infty$ && $\alpha =\beta =0$ & $0$ 
&&& EM$\Lambda$ \\ \hline
Barrow et al. (1986) & $+\infty$ & $0$ &
$\gamma =1$ & $\ne 0$ && $0$ & ED \\ \hline
Deser \& Mazur (1985) & $+\infty$ & $0$ &
$\alpha =\beta =0$ & $0$ &&& EM \\ \hline
Gott,III et al. (1986) & $+\infty$ & $0$ &
$\alpha =\beta =0$ & $0$ & $0$ && EM \\ \hline
Ba\~{n}ados et al. (1993) & $+\infty$ &&
$\alpha =\beta =0$ & $0$ && $0$ & E$\Lambda$ \\ 
\end{tabular}
\caption{Classification of solutions of three-dimensional Einstein 
equation. Abbreviations E, BI, M, D and $\Lambda$ stand for Einstein, 
Born-Infeld, Maxwell, dilaton and cosmological constant, respectively.}
\label{class}
\end{center}
\end{table}

\section{Solutions to the String Action}
\label{StringCase}
The case $n=1$, $\alpha=\beta=\gamma=4$ is particulary important,
since the solution with $n=1$ can be converted into the solution for the
string frame action via the conformal transformation
$g_{ij}\mapsto e^{-4\phi}g_{ij}$. 
We shall investigate this case in detail.
In this case, Eq.~(\ref{solution}) is expressed in terms of elementary
functions as
\begin{eqnarray}
&&f(r)=\left(\frac{r}{r_0}\right)
\left\{-M-\frac{2\Lambda r_1{}^2}{r_0}r
+\frac{4b^2r_1{}^2}{r_0}\left[1-\left(1+\frac{Q^2}{b^2r^2}\right)^{\half}
\right]r\right.\nonumber \\
&&\qquad\qquad\left.{}+\frac{4b|Q|r_1{}^2}{r_0}\ln\left[\left(1
+\frac{Q^2}{b^2r^2}\right)^{\half}+\frac{|Q|}{br}\right]\right\}.
\label{f_n=1}
\end{eqnarray}
The integration constant $M$ can be regarded as the total mass of the
system.
In fact, according to Brown and York \cite{BY,BCM}.
the (quasi-local) mass function $M(r)$ within the circle 
of radius $r$ becomes
\begin{equation}
M(r)=2f^{\half}\left(f_0^{\half}-f^{\half}\right)e^{-\delta},
\end{equation}
where $f_0=f_0(r)$ is a function defining the zero of mass
corresponding to some background space-time.
It will be natural to choose
\begin{equation}
f_0(r)=-\frac{2\Lambda r_1{}^2}{r_0{}^2}r^2,
\end{equation}
which implies that the Born-Infeld field vanishes in the background
space-time.
Then, $M$ is the total mass in the sense 
$M=\lim_{r\rightarrow+\infty}M(r)$.

In the following, we analyze the curvature singularity,
the causal structure and propagation of light front, where we show
some differences between the Born-Infeld and the 
Maxwell electrodynamics.

\subsection{Curvature Singularity}
There is a curvature singularity at the center $\{r=0\}$,
where the scalar curvature and the energy density diverge.
The asymptotic behavior near $\{r=0\}$ of these quantities
are shown in Table \ref{singular}.
In addition, the cases of the Einstein-Maxwell-dilaton-$\Lambda$ system
\cite{CM}, the Einstein-Maxwell-$\Lambda$ system (charged BTZ) \cite{BTZ},
the Einstein-Born-Infeld-$\Lambda$ system \cite{CG},
and the Einstein-dilaton-$\Lambda$ system ($Q=0$) are also shown
for comparison.
It can be seen that nonlinearity of
electrodynamics weakens the divergence of the curvature scalars
at least in the case of the dilaton-coupled systems.
The curvature scalars also diverge in the string frame.
The Table \ref{singular2} shows the behavior in the string frame.
\begin{table}[htbp]
\begin{center}
\begin{tabular}{ c c c c c}
System & $R$ & $\rho_{EM}$ & $\rho_{D}$ 
& $\rho_{\Lambda}$ \\ \hline\hline
EBID($Q\ne 0$) & $r^{-3}\ln r$ & $r^{-3}$ 
& $r^{-3}\ln r$ & $r^{-2}$\\ \hline
EMD($Q\ne 0$) & $r^{-4}$ & $r^{-4}$ & $r^{-4}$ & $r^{-2}$ \\ \hline
EBI($Q\ne 0$) & $r^{-1}$ & $r^{-1}$ & $0$ & $r^0$\\ \hline
charged BTZ & $r^{-2}$ & $r^{-2}$ & $0$ & $r^0$ \\ \hline
ED($Q=0$) & $r^{-3}$ & $0$ & $r^{-3}$ & $r^{-2}$ \\ 
\end{tabular}
\caption{Behavior of the scalar curvature and the energy densities
near the central singularity (Einstein frame).}
\label{singular}
\end{center}
\end{table}
\begin{table}[htbp]
\begin{center}
\begin{tabular}{ c c c c c}
System & $R$ & $\rho_{EM}$ & $\rho_{D}$
& $\rho_{\Lambda}$ \\ \hline\hline
EBID($Q\ne 0$) & $r^{-1}\ln r$ & $r^0$  & $r^{-1}\ln r$ & $r^0$   \\ \hline
EMD($Q\ne 0$) & $r^{-2}$ & $r^0$ & $r^{-2}$ & $r^0$  \\ \hline
ED($Q=0$) & $r^{-1}$ & $0$ & $r^{-1}$ & $r^0$  \\ 
\end{tabular}  
\caption{Same as in Table \protect\ref{singular}, but in the string frame.}
\label{singular2}
\end{center}
\end{table}

\subsection{Causal Structure}
A null hypersurface on which the horizon function $f(r)$ vanishes is a
Killing horizon.
The behavior of the function $f(r)$ determines the causal structure of the
space-time.
Introduce the following dimensionless quantities ($Q\ne0$),
\begin{equation}
x=\frac{b}{|Q|}r,~~q=\frac{2r_1Q}{r_0},
~~m=\frac{r_0M}{4br_1^2|Q|},
~~\lambda=\frac{\Lambda}{2b^2}.
\label{normalize}
\end{equation}
Then, Eq.~(\ref{f_n=1}) can be rewritten as
\begin{eqnarray}
F:=f/q^2=x\left\{-m+(1-\lambda)x-(1+x^2)^{\half}
{}+\ln\left[\frac{1+(1+x^2)^{\half}}{x}\right]\right\}.
\end{eqnarray}
It can be seen that the causal structure is determined by
essentially two parameters $m$ and $\lambda$.

For any $m$ and $\lambda$, the behavior of $F$ near $x=0$ is
\begin{equation}
F\sim -x\ln x ~(>0)~~(x\rightarrow +0).
\label{NearCenter}
\end{equation}
Using Eq.~(\ref{NearCenter}), 
we can see that the conformal radial coodinate
$r_\ast :=\int^r dr\sqrt{-g_{rr}/g_{tt}}$ 
remains finite as $x\rightarrow +0$, 
which implies the curvature singularity $\{r=0\}$ is timelike.
On the other hand, it can be seen that $r_{\ast}\rightarrow+\infty$
as $x\rightarrow+\infty$ for any $m$ and $\lambda$.
This implies that spatial infinity $\{r=\infty\}$ is a null
hypersurface.

The behavior of $F$ at spatial infinity
depends on $m$ and $\lambda$.
The causal structure is classified into several cases.\\
(i) $\lambda >0$ \\
In this case, $F\sim -\lambda x^2$ holds asymptotically
as $x\rightarrow\infty$ for any $m$.
Furthermore, $F=0$ has a simple positive root.
This implies that there exists a cosmological horizon.\\
(ii) $\lambda =0$ \\
The causal structure depends on $m$.
If $m$ is negative, there exists a cosmological horizon,
and the causal structure is similar to the $\lambda >0$ case.
When $m$ is positive or zero, $F$ is strictly posotive for any $x>0$.
Therefore, the central singularity is naked.\\
(iii) $\lambda <0$ \\
The causal structure depends on $m$.
The behavior of $F$ at spatial infinity is same as the $\lambda >0$
case except for the signature.
When $m$ satisfies
\begin{equation}
m=m_{\ast}(\lambda):=
\ln\left(1-\lambda+\sqrt{\lambda(\lambda-2)}\right),
\label{extremal}
\end{equation}
the equation $F=0$ has a multiple root.
Whether the central singularity is covered by the black hole
horizon depends on the value of $m$; \\
(iii-a) $m<m_{\ast}$ \\
In this case, $F>0$ for any $x>0$.
This implies that the central singularity is naked.\\
(iii-b) $m=m_{\ast}$ \\
This is the extreme case.
The black hole horizon is degenerated, and the causal structure is
similar to the four-dimensional extremal Reissner-Nordstr\"{o}m space-time.
The horizon radius becomes
$x=\left[\lambda (\lambda -2)\right]^{-\half}$.\\
(iii-c) $m>m_{\ast}$ \\
$F=0$ has two distinct positive roots.
These correspond the inner Cauchy horizon and the black hole horizon,
respectively.
The causal structure is similar to the four-dimensional nonextremal
Reissner-Nordstr\"{o}m space-time.

Here we show the effect of the nonlinear electrodynamics.
We rewrite Eq.~(\ref{extremal}) in terms of the original 
physical quantities.
Black hole horizons exist if and only if $\Lambda<0$ and
\begin{equation}
\frac{|Q|}{M}\leq\left(\frac{|Q|}{M}\right)_{\ast}:=
\frac{r_0}{4r_1^2b}\left(\ln\left[1-\frac{\Lambda}{2b^2}
+\sqrt{\frac{\Lambda}{2b^2}\left(\frac{\Lambda}{2b^2}-2\right)}
\right]\right)^{-1}.
\label{MtoQ1}
\end{equation}
To compare with the Einstein-Maxwell-dilaton system \cite{CM},
we take the Maxwell limit 
with fixing the other parameters;
\begin{equation}
\left(\frac{|Q|}{M}\right)_{\ast}\rightarrow
\frac{r_0}{4r_1^2}\left|\Lambda\right|^{-\half} \ \
( b\rightarrow\infty ).
\label{MtoQ2}
\end{equation}
For finite value of $b$, $\left(|Q|/M\right)_{\ast}$ 
is always larger than this limit value.
Therefore, there is a certain set of parameters $M$ and $Q$
for which black hole horizons exist in the Einstein-Born-Infeld-dilaton
system but do not exist in the Einstein-Maxwell-dilaton system.

\subsection{Propagation of Light Front}
Finally, we see the propagation of light front $\Sigma$, which is a
boundary of the region of disturbed electromagnetic field.
The motion of this characteristic surface $\Sigma$ can be investigated
by the method similar to \cite{NLSK}.
The electromagnetic field is discontinuous in the following sense;
\begin{equation}
\Bigl[\, F_{ij}\,\Bigr]_\Sigma=0,~~
\Bigl[\, \nabla_k F_{ij}\,\Bigr]_\Sigma=f_{ij}k_k,
\label{DiscontY}
\end{equation}
where $k_i$ is a vector normal to $\Sigma$.
The dilaton field $\phi$ and its first derivative 
are continuous at $\Sigma$
\begin{equation}
\Bigl[\, \phi\,\Bigr]_\Sigma=0,~~
\Bigl[\, \nabla_i\phi\,\Bigr]_\Sigma=0.
\end{equation}
We apply above conditions to the equation of motion (\ref{elemag}).
After short calculation,
we find that $k_i$ is not tangent to null geodesics of the background
space-time metric $g^{ij}$ but of the effective metric
\begin{equation}
g_{(eff)}^{ij}:= g^{ij}+4e^{-2\beta\phi}
\frac{\ddot{\cal L}}{\dot{\cal L}}F^i{}_k F^{jk}.
\label{EffMetric}
\end{equation}
Using Eqs.~(\ref{BI_Lag}), (\ref{metric}), (\ref{strength}) and (\ref{elec}),
the effective metric can be rewritten as
\footnote{Covariant components of the effective metric are defined as
$g_{i\ell}^{(eff)}g_{(eff)}^{\ell j}=\delta_i^j$. 
See Ref.\cite{NLSK}.}
\begin{equation}
g_{(eff)}=h^{-1}\left(-fdt^2+\frac{e^{2\delta}}{f}dr^2\right)
+r^2d\varphi^2.
\label{EffMetric2}
\end{equation}
A generator of the null geodesic of the effective metric
(\ref{EffMetric2}) is written as $k^i=\d x^i/\d\tau$.
Then, $k^i$ satisfies following equations,
\begin{eqnarray}
&& \left(\frac{\d r}{\d\tau}\right)^2+V(r)=0, \nonumber\\
&& V(r):= \frac{e^{-2\delta}}{h^2}\left(\frac{hf}{r^2}L^2-E^2\right),
\label{EffecPoten}
\end{eqnarray}
where $E$ and $L$ are conserved quantities defined by
\begin{equation}
E=h^{-1}f\frac{\d t}{\d\tau},~~
L=r^2\frac{\d\varphi}{\d\tau}.
\end{equation}
The light front can exist in the region $V<0$.
In the following, we consider the case $n=1$.
The behavior of potential $V$ is determined by three parameters
$m$, $\lambda$ and
\begin{equation}
w:=\left(\frac{r_0}{2br_1}\right)^2\left(\frac{E}{L}\right)^2.
\end{equation}

\begin{figure}
\centerline{\epsfxsize=7cm \epsfbox{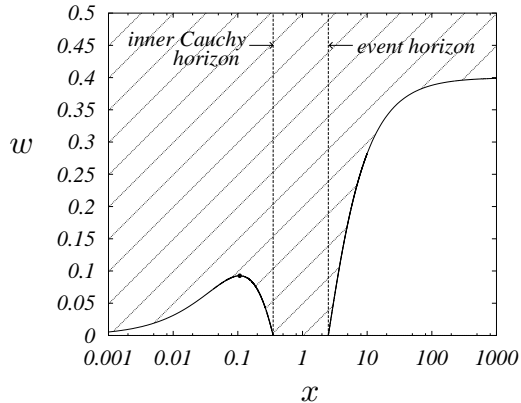}}
\caption{Allowed region of light fronts (shaded region) is shown
in the $x-w$ plane for a fixed background space-time
in the case of Einstein-Born-Infeld-dilaton system
($\lambda=-0.4$, $m=1.2$).
The filled circle corresponds to the unstable circular orbit.}
\label{LightBINS}
\end{figure}
\begin{figure}
\centerline{\epsfxsize=7cm \epsfbox{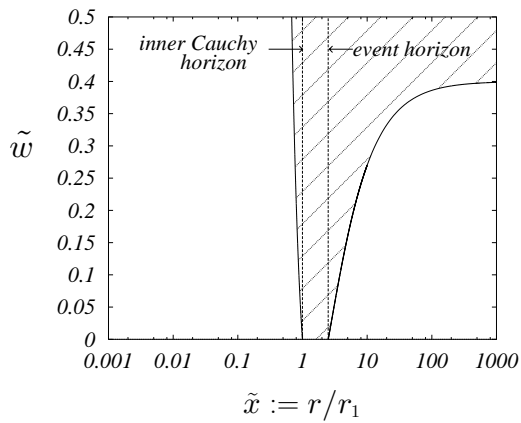}}
\caption{Same as in Fig.\protect\ref{LightBINS},
but in the case of Einstein-Maxwell-dilaton system 
($\tilde{\lambda}=-0.4$, $\tilde{m}=1.4$).}
\label{LightMNS}
\end{figure}

The Figure \ref{LightBINS} shows the region $V<0$.
The background parameters $m$ and $\lambda$ are fixed and satisfy
$\lambda<0$ and $m>m_\ast$, so that 
the black hole horizon and the inner Cauchy horizon exist.
For comparizon, we also show a same calculation 
in the case of the Maxwell limit.
When ${\cal L}=-x$, the second term on the right hand side of 
Eq.~(\ref{EffMetric}) vanishes.
Therefore, $k_i$ is tangent to null geodesics of the background space-time,
which is well-known result.
Inserting $h=1$ and Eq.~(\ref{CMa}) with $n=1$ into 
Eq.~(\ref{EffecPoten}), we find that the potential function $V$ is 
determined by the following three parameters
\begin{equation} 
\tilde{m}=\frac{r_0M}{2r_1Q^2},~~
\tilde{\lambda}=\frac{r_1^2}{Q^2}\Lambda,~~
\tilde{w}=\frac{1}{2}\left(\frac{r_0}{Q}\right)^2\left(
\frac{E}{L}\right)^2.
\label{Norm3}
\end{equation}
The Figure \ref{LightMNS} shows the region $V<0$ with fixed background
parameters $\tilde{m}$ and $\tilde{\lambda}$
for which no Killing horizons exist.

We can see some differences between the Born-Infeld and the Maxwell 
electrodynamics.
Let us consider the light fronts coming from infinity to the 
center $\{r=0\}$.
They can arrive at the center in the Born-Infeld case,
while they are scattered in the Maxwell case if $L\ne 0$.
Furthermore, the unstable circular orbit exist only in
the Born-Infeld case.
These differences between the Born-Infeld and the Maxwell
electrodynamics can also be seen when the naked singularity
or the cosmological horizon exist.

\section{Conclusion}
\label{Conclusion}
We have given a seven parameter family of static, circularly symmetric, 
and pure electrically charged solutions of the three-dimensional 
Einstein-Born-Infeld-dilaton system.
The electric field remains finite near the central singularity,
which shows a nonlinear effect of the Born-Infeld field.
The solution reduces to known black hole solutions of various
theories by appropriately taking limits of parameters.

We have studied the low energy string case ($n=1$) in detail.
The solution is essentially described by the mass parameter 
$m$ and the cosmological constant parameter $\lambda$.
The critical value of mass parameter $m_\ast$
exists for given $\lambda$, 
such that the black hole horizons exist if and only if 
$\lambda <0$ and $m\ge m_{\ast}$ \cite{Ida}.
The causal structure of charged black hole solutions ($\lambda <0$ case)
is similar to those of the four-dimensional Reissner-Nordstr\"{o}m solution.
The spatial infinity is not a timelike but a null hypersurface.
This is an effect of the dilaton field \cite{CM}.

We have revealed some differences between the Born-Infeld nonlinear
electrodynamics and the ordinary Maxwell electrodynamics.
(i) The curvature scalar and the energy density diverge
as $r\rightarrow 0$ in both the Einstein frame and the string frame.
In the dilaton-coupled system,
the divergence of curvature scalars is weaker in the 
Born-Infeld case than in the Maxwell case.
(ii) The lower bound of mass for which black hole horizons exist is
lower in the Born-Infeld case,
so that it is easier to form a black hole horizon
in the Born-Infeld case
in the sense that the parameter region corresponding to the 
black holes is wider.
(iii) We have considered the light fronts coming from infinity towards 
the center with non-zero impact parameter.
The light fronts are always scattered in the Maxwell electrodynamics,
while they can reach the center in the Born-Infeld electrodynamics.

\section*{Acknowledgments}
We would like to acknowledge many helpful discussions with Professor
Humitaka Sato and Prof. Takeshi Chiba.
D.I. was supported by JSPS Research and this research was
supported in part by the Grant-in-Aid for Scientific Research 
Fund (No.~4318).

\end{document}